\documentclass[usenatbib,fleqn,useAMS]{mnras}

\usepackage{graphicx}
\usepackage{amsmath}
\usepackage{mathrsfs}
\usepackage[varg]{txfonts}
\usepackage{braket}
\usepackage{cancel}
\usepackage{color}

\renewcommand\le\oldleq
\renewcommand\ge\oldgeq

\renewcommand\pi\upi
\newcommand\Atan{\mathrm{arctan}}

\title[A Halo Substructure in Gaia Data Release 1]
{A Halo Substructure in Gaia Data Release 1}

\author[Myeong, Evans, Belokurov, Koposov \& Sanders]
       {G.~C.~Myeong$^1$\thanks{E-mail:~gm564,nwe,vasily,koposov,jls@ast.cam.ac.uk},
         N.~W.~Evans$^1$, V.~Belokurov$^1$, S.E.~Koposov$^{1,2}$ \& J.L.~Sanders$^1$ \\$^1$Institute of
         Astronomy, University of Cambridge, Madingley Road, Cambridge
         CB3~0HA.
	\\$^2$ McWilliams Center for Cosmology, Department of Physics, Carnegie
Mellon University, 5000 Forbes Avenue, Pittsburgh, PA 15213, USA
	}

\pagerange{\pageref{firstpage}--\pageref{lastpage}}\volume{000}\pubyear{2017}
\date{version \today.}

\begin{document}
\label{firstpage}
\maketitle

\begin{abstract}
We identify a halo substructure in the Tycho Gaia Astrometric Solution
(TGAS) dataset, cross-matched with the RAVE-on data release. After
quality cuts, the stars with large radial action ($J_R > 800$
kms$^{-1}$ kpc) are extracted. A subset of these stars is clustered in
longitude and velocity and can be selected with further cuts. The 14
stars are centered on $(X,Y,Z) \approx (9.0,-1.0,-0.6)$ kpc and form a
coherently moving structure in the halo with median $(v_R,v_\phi,v_z)
= (167.33,0.86,-94.85)$ kms$^{-1}$.  They are all metal-poor giants
with median [Fe/H] $=-0.83$. To guard against the effects of distance
errors, we compute spectrophotometric distances for the 8 out of 14
stars where this is possible. We find that 6 of the stars are still
comoving. These 6 stars also have a much tighter [Fe/H] distribution
$\sim -0.7$ with one exception ([Fe/H] = -2.12). We conclude that the
existence of the comoving cluster is stable against changes in
distance estimation and conjecture that this is the dissolving remnant
of a long-ago accreted globular cluster.
\end{abstract}

\section{Introduction}

Over the last decade, many streams have been identified in the stellar
halo of the Milky Way, usually as overdensities of main-sequence
turn-off stars in resolved star maps from wide area photometric
surveys~\citep[see e.g.,][]{Be06,Gr09,Ne15}.  An alternative method is
to identify substructure kinematically as samples of stars with
similar chemistry moving in a distinct and coherent way. Though less
widely used, this has had some striking successes, including the
famous identification of the Sagittarius dwarf~\citep{Ib94}, the halo
stream found by \citet{He99} in {\it Hipparcos} data, and the globular
cluster streams found by \citet{Sm09} in Sloan Digital Survey Stripe
82 data.

The first data releases (DR1) of the Gaia satellite~\citep{GA1,GA2}
provides us with a new vista of the Solar neighbourhood. The primary
astrometric catalogue in Gaia DR1 is TGAS (Tycho Gaia Astrometric
solution), which uses data from Tycho-2~\citep{Ho00} to provide a $30$
year baseline for astrometric calculations. It has $2\,057\,050$
entries. When cross-matched with RAVE-on~\citep{Ca16,Ku17}, this
gives a catalogue of $180\,929$ stars with full six-dimensional phase
space information, as well as associated stellar spectral
quantities. TGAS cross-matches are also possible with
LAMOST~\citep{Lu15} and APOGEE~\citep{An14}, though they are somewhat
smaller in size with $78\,579$ and $12\,061$ entries respectively. Not all
the entries in the three cross-matched catalogues are distinct. When
stars overlap in the surveys, we take the data with the smallest
relative error in radial velocity. This yields a final master
catalogue of $268\,588$ stars where the RAVE-on contribution is $180\,454$
stars.  The master catalogue is a natural arena in which to search for
kinematic substructure.

This {\it Letter} identifies a group of stars moving on strongly
radial orbits in the TGAS cross-matched master catalogue. The stars
have similar metallicities, and the simplest explanation of their
unusual kinematics is that they are the residue of an ancient halo
structure. Section 2 describes our treatment of the data and
extraction of the substructure stars. Section 3 discusses the selected
stars, together with estimates for the age and future evolution of the
substructure.

\begin{figure*}
\begin{center}
\includegraphics[width=0.9\textwidth]{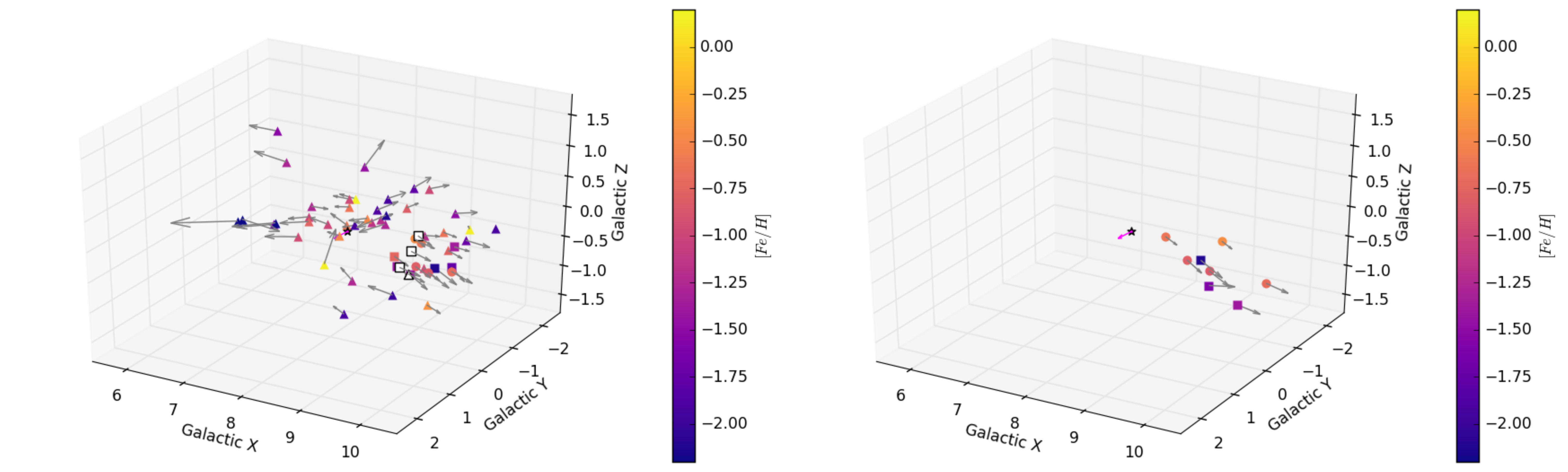}
\end{center}
\caption{Left: The spatial structure of the $56$ stars
    satisfying the quality cuts and with $J_R> 800$ kms$^{-1}$ kpc.
    Stars are colour-coded according to metallicity if known, whilst
    the arrow indicates the magnitude and direction of the spatial
    velocity.  There is a clear clustering of stars at $(X,Y,Z)
    \approx (9.0,-1.0,-0.6)$ kpc, which have radial motion dominating
    their total velocity. The $14$ stars that may belong to the
    comoving cluster are shown as circles or squares, while the
    remainder are shown as triangles. Right: Using spectrophotometric
    distances, the $14$ candidate members are re-examined. $5$ stars
    (circles) are retained as confirmed members as both their
    distances and metallicities are similar. Of the remaining $9$
    stars, $6$ do not have spectrophotometric distances. However, $2$
    are not comoving according to the new distances, whilst $1$ is
    comoving but has markedly different metallicity. These $3$ objects
    are marked with squares. Finally, the Sun is marked on both plots
    as a star.}
\label{fig:threed}
\end{figure*}
\begin{figure*}
\begin{center}
\includegraphics[width=0.9\textwidth]{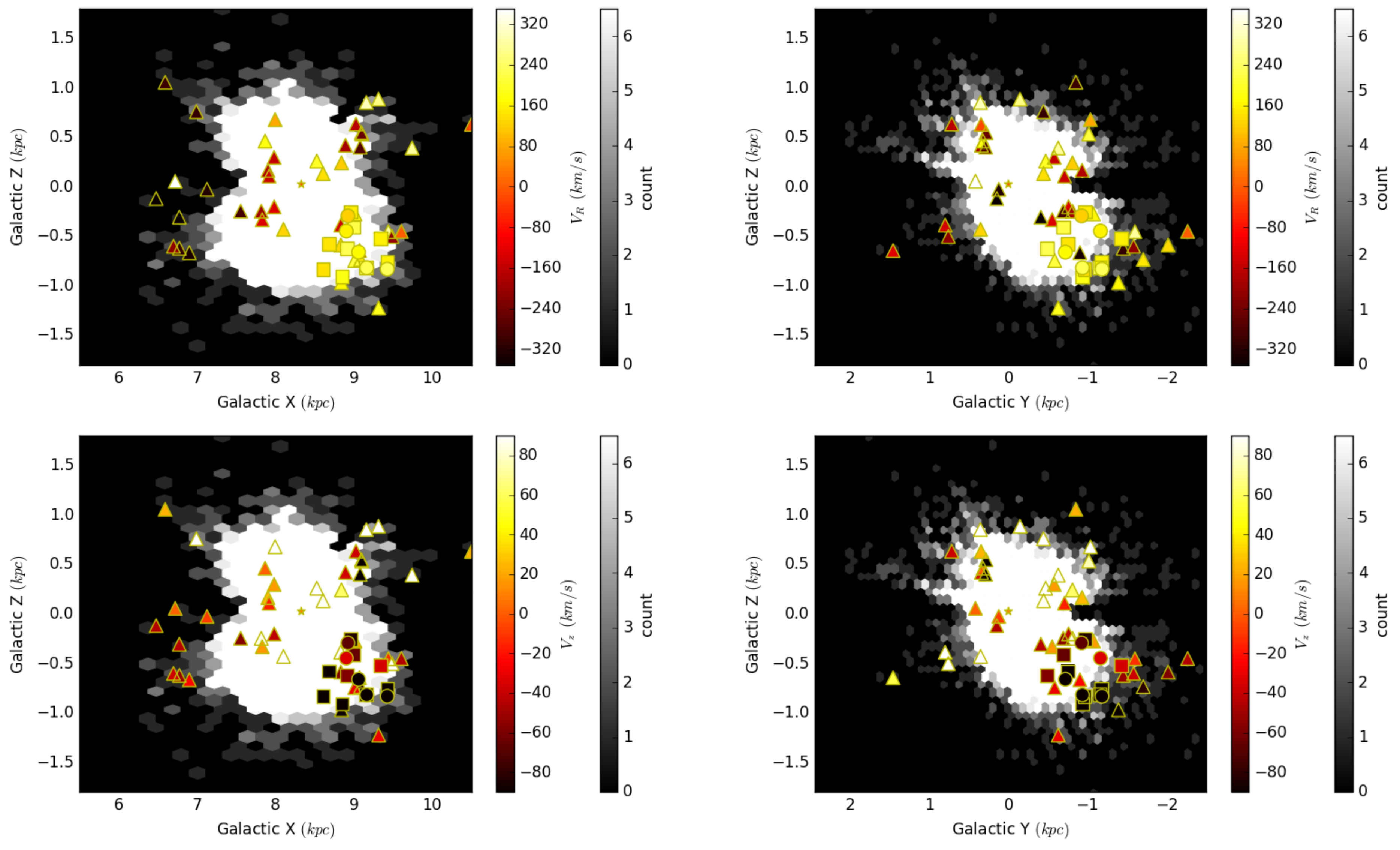}
\end{center}
\caption{Locations of the $56$ high-quality stars with $J_R> 800$
  kms$^{-1}$ kpc in the Galactic ($X,Y$) and ($Y,Z$) planes using the
  \citet{As16} distances. Circles (5 confirmed members) and squares (9
  possibles) show stars associated with the comoving clump, whilst
  triangles show the rejected stars. Symbols are coloured according to
  the radial velocity (upper panels) or vertical velocity (lower
  panels). The colours of the symbols show that the stars of interest have
  quite similar velocitiues, both vertical and radial.}
\label{fig:xydist}
\end{figure*}
\begin{figure}
\begin{center}
\includegraphics[width=0.5\textwidth]{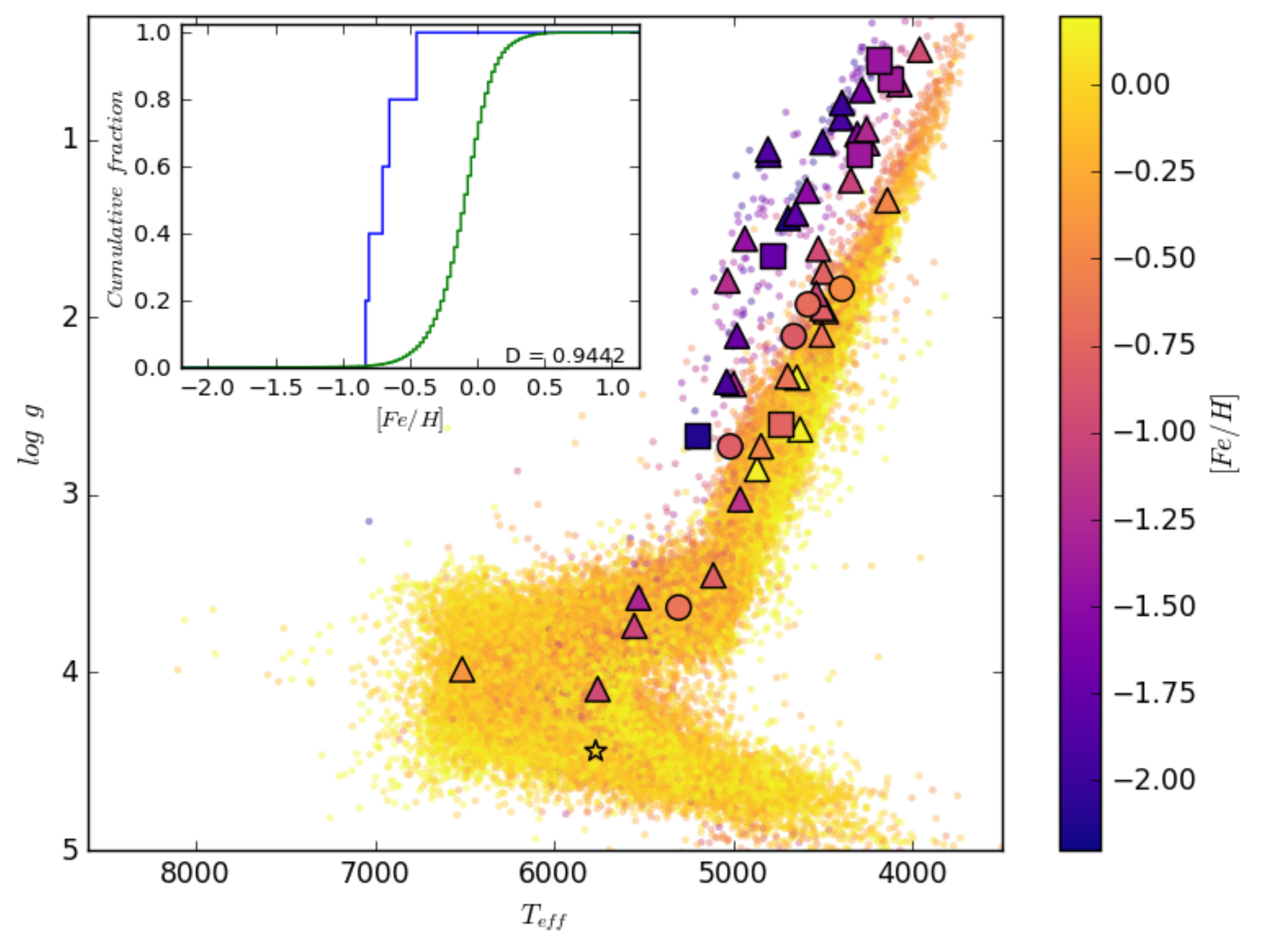}
\end{center}
\caption{Surface gravity versus effective temperature for all stars in
  the master catalogue with spectroscopic parameters. The $11$
  comoving stars with known metallicity are shown as circles
  (confirmed members) or squares (possibles), the remainder as
  triangles. The location of the Sun is shown as a star for
  reference. The inset shows the normalised cumulative histogram of
  [Fe/H] for the 5 secure comoving stars (in blue) and for the thin
  disk (in green). The $D$ value for the KS-test indicates that the
  two distributions are very different.}
\label{fig:metals}
\end{figure}
\begin{table}
\begin{center}
  \begin{tabular}{lrc}
\hline
\hline
\multicolumn{1}{c}{RAVE ID} &
\multicolumn{1}{c}{RA} & \multicolumn{1}{c}{DEC}\\
\hline
J063314.9-284345 & 98.312 & -28.729\\
J062653.3-355032 & 96.722 & -35.842\\
J052906.3-240844 & 82.276 & -24.145\\
J050209.4-235127 & 75.539 & -23.858\\
J050519.6-264950 & 76.332 & -26.831\\
\hline
  \end{tabular}
\end{center}
  \caption{RAVE identifiers, as well as right ascension and declination, for the 5 confirmed members.}
  \label{table:stars}
\end{table}

\begin{table*}
\begin{center}
  \begin{tabular}{lrrrrrrrrc}
\hline
\hline
\multicolumn{1}{c}\null &
\multicolumn{1}{c}{[Fe/H]} & \multicolumn{1}{c}{$\log g$} &
\multicolumn{1}{c}{$T_{\rm eff}$} &
\multicolumn{1}{c}{$(X,Y,Z)$} & \multicolumn{1}{c}{$(v_R,v_\phi,v_z)$} & \multicolumn{1}{c}{$e$} &
\multicolumn{1}{c}{$R_{\rm apo}$} &
\multicolumn{1}{c}{$R_{\rm peri}$} &
\multicolumn{1}{c}{$z_{\rm max}$}
\\
\multicolumn{1}{c}\null &
 \multicolumn{1}{c}{\null} & \multicolumn{1}{c}{\null} &
 \multicolumn{1}{c}{(K)} & \multicolumn{1}{c}{(kpc)} &
 \multicolumn{1}{c}{(kms$^{-1}$)} & \multicolumn{1}{c}{\null} &
 \multicolumn{1}{c}{(kpc)} &
 \multicolumn{1}{c}{(kpc)} &
  \multicolumn{1}{c}{(kpc)}
\\
\hline
Median&      -0.83    &    1.93    &    4662    &    (9.00,-0.94,-0.62)    &
      (167.33,0.86,-94.85)    &    0.940     & 12.79 & 0.40
      & 3.50 \\
Median Absolute Deviation&      0.28     &    0.74    &    355    &    (0.15,0.21,0.18)    &
      (21.47,10.18,30.67)    &   0.017   & 1.55  & 0.11 &
      1.45 \\
      \hline
Median&      -0.70    &    2.11    &    4662    &    (9.06,-0.94,-0.66)    &
      (167.33,8.79,-94.86)    &    0.948     & 12.31 & 0.40
& 3.50 \\
\null &      \null    &    \null    &    \null    &    (9.13,-0.90,-0.53)    &
      (168.57,11.24,-72.96)    &    0.947     & 12.02 & 0.29
& 4.50 \\
Median Absolute Deviation&      0.10     &    0.27    &    268    &    (0.14,0.21,0.17)    &
      (44.35,8.11,27.22)    &   0.011   & 1.98  & 0.13 &
1.69 \\
\null &      \null     &    \null    &    \null    &    (0.21,0.37,0.25)    &
      (52.23,1.38,7.13)    &   0.006   & 1.92  & 0.08 &
2.65 \\
\hline
  \end{tabular}
\end{center}
  \caption{Upper: The medians and median absolute
      deviations in spectroscopic and orbital properties of all 14
      possible stars in the comoving group. Lower: The same, but for
      the subset of 5 stars that are unambiguous members. In this
      case, the velocities and orbital parameters are given first
      using \citet{As16} distances, and then using spectrophotometric
      distances.}
  \label{table:params}
\end{table*}

\section{Extraction of the Member Stars}

First, the proper motions are converted to velocities using the
unbiased, inferred distance estimates of \citet{As16}. To ensure a
high quality sample, we impose a cut that the error in the radial
velocity $\epsilon_{\rm RV} < 10$ kms$^{-1}$, the error in the total
velocity $\epsilon_{\rm Vtot} < 25$ kms$^{-1}$ and the relative error
in the total velocity is $< 10 \%$.  This reduces the sample to
$73\,268$ stars ($66\,891$ using the RAVE-on measurements, the
remainder from LAMOST and APOGEE). We now use the Galactic potential
{\tt MWPotential2014} in \citet{Bo15} to compute the radial and
azimuthal actions \citep[$J_R$ and $J_\phi$, using the adiabatic
  approximation e.g.][]{Binney2010}. The bulk of the stars are moving
on nearly circular orbits in the thin disk. Nonetheless, there are
extensions of stars in the high action regime, which are predominantly
metal-poor and are moving on eccentric orbits.

We retain only stars with a radial action $J_R > 800$ kms$^{-1}$ kpc.
This cut leave us with $56$ stars moving on predominantly eccentric
orbits ($41$ are from the RAVE-on crossmatch, $15$ from the other
crossmatches).  Fig.~\ref{fig:threed} shows the spatial distribution
of all these stars, colour-coded according to metallicity (if known),
with arrows representing their velocity vectors. The existence of a
comoving cohort of stars at $(X,Y,Z) \approx (9.0,-1.0,-0.6)$ kpc is
evident. This is confirmed by a visual inspection of histograms of
velocity components resolved with respect to the cylindrical polar
coordinate ($v_R,v_\phi,v_z$), which betrays clear peaks corresponding
to the comoving clump. To formalize their extraction, we perform
$2.5\sigma$ clipping based on $v_R,v_\phi,v_z$ and orientation angle
$\Psi$, individually. Here, $\Psi = {\Atan}(-V/U)$, where $U$ and $V$
are the Galactocentric Cartesian velocity components along the $X$ and
$Y$ directions. So, $\Psi$ is the angle of star's motion in Galactic
plane. Then, only those stars falling within the range of the median
$\pm 2.5\sigma$ (measured after the clipping) are retained. This
process is repeated until no further stars are rejected. This leaves a
set of $14$ stars possibly belonging to a comoving clump with strongly
radial orbits. They are shown as circles or squares in the left panel
of Fig.~\ref{fig:threed}, color-coded according to metallicity.  All
these stars lie in the TGAS and RAVE-on crossmatch.

For this set of 14 stars, we measured the median and $\sigma$
(standard deviation) in each ($U,V,W$) component. To ensure that no
possible members have been missed, we returned to the original sample
of $73\,268$ and searched for all stars in the ($U,V,W$) velocity box
bounded by the median $\pm 2\sigma$ in each velocity component.  This
covers a comparable but wider range then the `minimum to maximum'
range in each velocity component of the 14 initial stars, so it is
wide enough to contain our initial sample and any others that may have
some comparable motion. However, no further stars are identified as
possible members.

\citet{As16} used a prior suitable for nearby disk stars and they
caution that their distances may be underestimates when objects lie
beyond 2 kpc from the Sun. Our comoving candidates have heliocentric
distance $\approx 1.8$ kpc, so it is prudent to cross-check our
candidates with another different distance estimator.  There are 8
RAVE-on stars among our candidates that satisfy recommendations
advocated by \citet{Ca16}, namely they have $\texttt{teff\_sparv} < $
8000 K, the RAVE spectral parameters $c_1,c_2,c_3$ reported as `n' or
normal, and a reduced $\chi^2 < 3$ as recorded by `the Cannon'
pipeline. For these stars, we can use the technique pioneered by
\citet{Bu10} to compute the spectrophotometric distance distribution
folded with the TGAS parallax distribution for each
star. Specifically, we use the PARSEC isochrones \citep{Bressan2012}
and the extinction map of \citet{Gr15} to obtain the probability
distribution function of the distance given the spectroscopic
parameters ($T_\mathrm{eff}$, $\log g$, [Fe/H]), apparent 2MASS
magnitudes ($J$, $H$ and $K_s$) and the TGAS parallaxes along with
their associated uncertainties (or covariances where available). We
adopt an identical prior to that used in \cite{Binney2014}. The right
panel of Fig.~\ref{fig:threed} shows the new view of the comoving
candidates, Of the 8 stars with spectrophotometric distances, 5 have a
common velocity and metallicity. We regard these as confirmed members
and they are shown as circles. Their RAVE identifiers, as well as
right ascensions and declination, are given in
Table~\ref{table:stars}. There are 3 remaining stars, 2 of which are
no longer comoving, and 1 of which is comoving but has a discrepant
metallicity. We regard these as merely possible members, pending
confirmation of the distance or metallicity, and they are shown as
squares. In other words, we have 5 confirmed members and 9 possible
members, including the stars without spectrophotometric distances.

In the panels of Fig.~\ref{fig:xydist}, the circles and squares show
our selected clump or co-moving group, while the triangles are the
rejected stars (but still possessing $\epsilon_{\rm RV} < 10$
kms$^{-1}$, $\epsilon_{\rm Vtot} < 25$ kms$^{-1}$, relative velocity
error $< 10 \%$ and $J_R > 800$ kms$^{-1}$ kpc). The stars are
colour-coded according to their radial and vertical velocities, so a
coherent group stands out as a clump of objects with similar
colouring.  The significance of the comoving group above the
background level can be measured in Galactocentric $(X,Y)$, $(X,Z)$
and $(Y,Z)$ planes. The number of stars (with $\epsilon_{\rm RV} < 10$
kms$^{-1}$, $\epsilon_{\rm Vtot} < 25$ kms$^{-1}$, relative velocity
error $< 10 \%$ and $J_R > 800$ kms$^{-1}$ kpc) in the comoving group
region (drawn as an aperture that encloses the group) was compared
with the number of stars in other regions (representing the background
level) across the plane. The significance of the comoving group above
the background level is $3.2\sigma$, $7.2\sigma$ and $4.8\sigma$ for
each Galactocentric $(X,Y)$, $(X,Z)$ and $(Y,Z)$ planes respectively.

The comoving sample of $14$ stars has median position $(X,Y,Z) =
(9.00, -0.94, -0.62)$ kpc and velocity $(v_R,v_\phi,v_z) =
(167.33,0.86,-94.85)$ kms$^{-1}$ in the Galactic rest frame. By
integrating orbits in potential {\tt MWPotential2014} from
\citet{Bo15}, we obtain a median apocentric distance of $\sim 13$ kpc
and pericentric distance of $\sim 0.4$ kpc for stars in the clump.
The high eccentricity (median $e = 0.940$) and low apocentric
distances suggest that the clump may be the relic of an object that
fell into the halo long ago and whose apocentric distance has been
reduced by dynamical friction over a number of pericentric
passages. These values are recorded in Table~\ref{table:params}. where
we also give the corresponding figures if the sample is restricted to
just the 5 stars with confirmed membership.

\begin{figure}
\begin{center}
\includegraphics[width=0.5\textwidth]{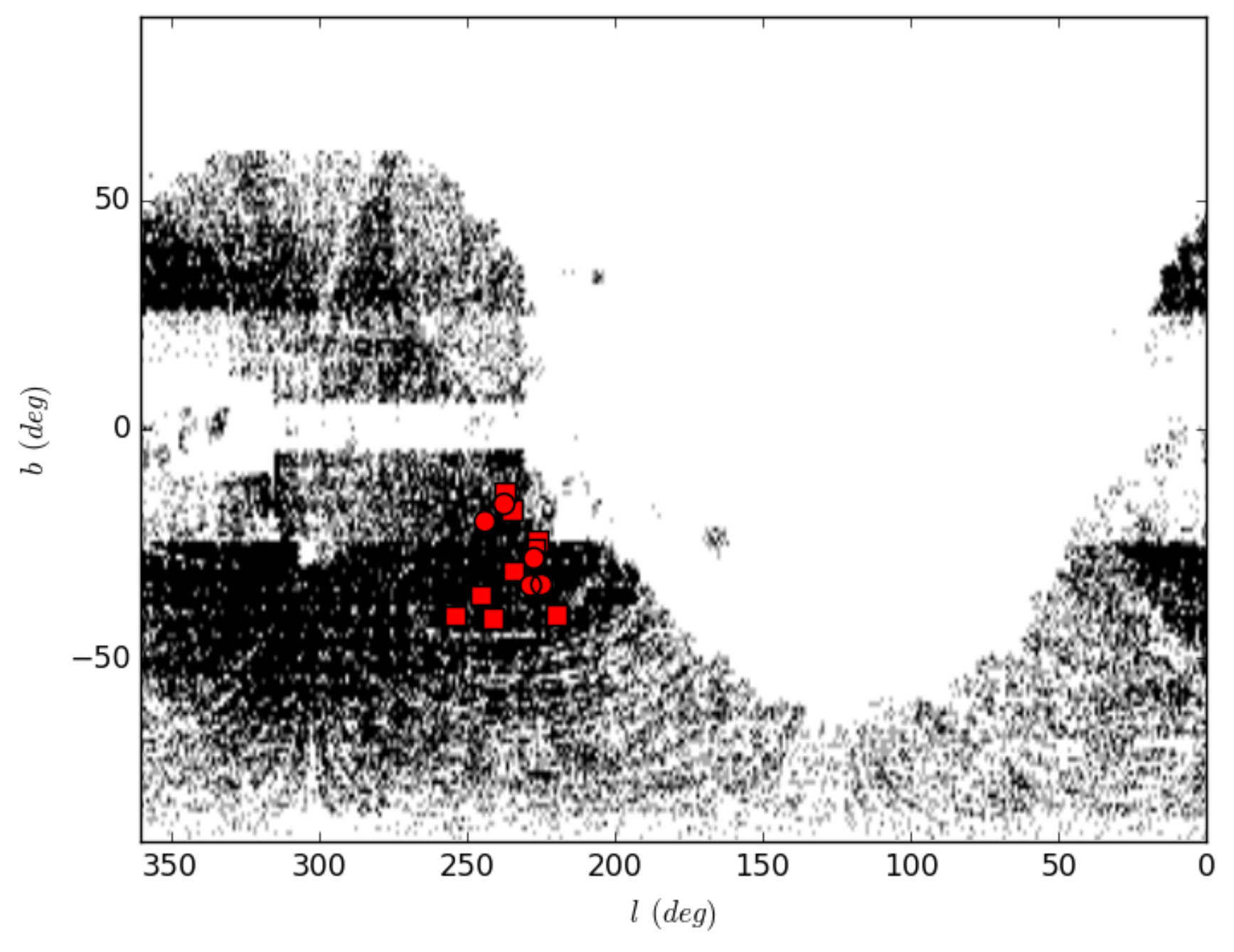}
\end{center}
\caption{The RAVE footprint in Galactic coordinates.  The 14 comoving
  clump stars are shown as circles (confirmed members) or squares
  (possibles). As the edge of the footprint abuts the clump, there may
  be an extension of the substructure that is currently missing.}
\label{fig:rave}
\end{figure}

Fig.~\ref{fig:metals} shows the spectroscopic properties of the
stars. They are all metal-poor giants with median [Fe/H] $= -0.83$. By
conducting a Kolmogorov-Smirnov test, we can verify that the
metallicities are not consistent with being drawn from the thin disk
($D=0.944$). This is illustrated in the inset to the figure.  The
medians and median absolute deviations in the spectroscopic quantities
are also noted in Table~\ref{table:params}.  As the stars in the
comoving clump are exclusively drawn from the RAVE survey, an
immediate concern is that the footprint of the survey may affect the
results. The RAVE footprint and the comoving clump are shown in
Galactic coordinates in Fig.~\ref{fig:rave}. The positions of the 14
stars are marked in red. The distribution of circles (confirmed
members) is suggestive of a stream moving from upper left to lower
right. The three squares (possibles) at ($\ell \approx 245^\circ,
b\approx -40^\circ$) then lie $\sim 300$ pc off the stream. This also
hints that these three stars are less likely to be members of the
structure. However, this is not conclusive as it is conceivable that a
globular cluster stream could broaden over time. Note that the
substructure does extend towards the edge of the RAVE footprint, and
so it is conceivable that it may be larger than is apparent from the
TGAS and RAVE cross-match.

\section{Discussion and Conclusions}

Using crossmatches of the Tycho Gaia Astrometric Solution (TGAS) with
the Radial Velocity Experiment (RAVE-on), we have extracted a sample
of 14 stars with unusual kinematics. The stars are all moving on
strongly radial orbits and are clumped at $(X,Y,Z) \approx
(9.0,-1.0,-0.6)$ kpc. The median eccentricity of the stars is $0.940$.

We checked that this comoving group is resilient against distance
errors. For 8 out of the 14 stars, it is possible to calculate
spectrophotometric distances using the methods of \citet{Bu10}.  6 of
the stars still show a comoving trend. These 6 stars also have a much
tighter [Fe/H] distribution $\sim -0.7$ with one exception ([Fe/H] =
-2.12). The median absolute deviation of [Fe/H] is $0.10$.  So, the
motion, spatial location and metallicity of at least 5 stars are well
confined in both distance estimates, suggesting that they are comoving
and have a common origin.

The simplest interpretation of this coherently moving substructure is
that it is the residue of a long past accretion event. The small
metallicity spread in the 5 secure members suggests that the
progenitor was more likely to be a globular cluster than a dwarf
galaxy. The comoving group may be identified with the nucleus of this
disintegrating body. Naturally, we would expect tidal streams also to
be present, but in an old accretion event, the surface density of
stream material mey be very low.  Although some streams can be caused
by resonance effects, such as the Hercules Stream~\citep{De00} and
some can be caused by interactions with perturbers, such as the
Aquarius Stream~\citep{Ca14}, neither option seems likely
here. Resonance effects usually give structures of modest
eccentricity, as nearly circular orbits can usually couple most easily
to the perturbation.

The large globular cluster $\omega$Centauri has long been thought to
be the remnant nucleus of an accreted dwarf galaxy \citep{Fr93}, as it
shows evidence of multiple stellar populations with spreads in the age
and metallicity \citep[e.g.][]{Vi07}. If our comoving group is
associated with $\omega$Centauri, for example, by sharing a common
progenitor, some clues may be found from stellar parameters or
dynamics. RAVE DR4 \citep{Ko13} provides an estimate of the age for 14
comoving group members. The median age is $9.8$ Gyr with standard
deviation of $1.2$ Gyr. The age-metallicity relation for
$\omega$Centauri presented by \citet{Vi14} shows some stars with
metallicity and the age comparable to our comoving group stars, even
though this estimated age is a relative age.  Yet, the orbit of the
comoving group appears to be highly eccentric with $v_\phi$ close to
zero.  It is ambiguous as to whether the group is prograde or
retrograde, while $\omega$Centauri clearly has a less eccentric and
retrograde orbit \citep{Ma00}.  In addition, the estimated values of
energy and angular momentum of the comoving group stars do not show a
clear link with $\omega$Centauri, suggesting that this comoving group
is very unlikely to be assocaited with it.

Although there have already been searches for halo overdensities in
the TGAS-RAVE cross-match by~\citet[][herafter H17]{He17}, the
substructure identified in this paper appears to be new. It does not
correspond to any of the overdensities labelled VelHel-1 to 9 in
H17. As the underlying Galactic potential is different between the two
papers, the estimated values of energy $E$ of the stars are different.
H17 either convert parallaxes to distances via the reciprocal (70 \%
of their sample) or use the RAVE distances (30 \%). These are not the
same as our spectrophotometric distances, so the azimuthal actions
$J_\phi$ of the stars are also different.  This affects the integrity
of the substructures identified in H17, which we find to be indistinct
and smeared out. As judged by the Tycho-IDs, there are only 4 stars in
our substructure that overlap with H17, two in the VelHel-1 and 2 in
VelHel-8. The claimed enhancements in H17 therefore do not appear to
be related to our substructure, which is a coherent entity in
configuration and velocity space.

\section*{acknowledgements}
GCM thanks Boustany Foundation, Cambridge Commonwealth, European \&
International Trust and Issac Newton Studentship for their support on
his work. SK thanks the United Kingdom Science and Technology Council
(STFC) for the award of Ernest Rutherford fellowship (grant number
ST/N004493/1). We thank the referee (Carl Grillmair) and Paul McMillan
for helpful comments on the manuscript. This work has made use of data
from the European Space Agency (ESA) mission Gaia
(https://www.cosmos.esa.int/gaia), processed by the Gaia Data
Processing and Analysis Consortium (DPAC,
https://www.cosmos.esa.int/web/gaia/dpac/consortium). Funding for the
DPAC has been provided by national institutions, in particular the
institutions participating in the Gaia Multilateral Agreement.

\label{lastpage}

\begin{thebibliography}{}

\bibitem[Anders et al.(2014)]{An14} Anders, F., Chiappini, C.,
  Santiago, B.~X., et al.\ 2014, \aap, 564, A115

\bibitem[Astraatmadja \& Bailer-Jones(2016)]{As16} Astraatmadja,
  T.~L., \& Bailer-Jones, C.~A.~L.\ 2016, \apj, 833, 119

\bibitem[Belokurov et al.(2006)]{Be06} Belokurov, V., Zucker, D.~B.,
  Evans, N.~W., et al.\ 2006, \apjl, 642, L137

\bibitem[\protect\citeauthoryear{Binney}{2010}]{Binney2010} Binney J., 2010, MNRAS, 401, 2318

\bibitem[\protect\citeauthoryear{Binney et al.}{2014}]{Binney2014} Binney J., et al., 2014, MNRAS, 437, 351

\bibitem[Bovy(2015)]{Bo15} Bovy, J.\ 2015, \apjs, 216, 29

\bibitem[\protect\citeauthoryear{Bressan et al.}{2012}]{Bressan2012} Bressan A., Marigo P., Girardi L., Salasnich B., Dal Cero C., Rubele S., Nanni A., 2012, MNRAS, 427, 127

\bibitem[Burnett \& Binney(2010)]{Bu10} Burnett, B., \& Binney,
  J.\ 2010, \mnras, 407, 339

\bibitem[Casey et al.(2014)]{Ca14} Casey, A.~R., Keller, S.~C., Alves-Brito, A., et al.\ 2014, \mnras, 443, 828

\bibitem[Casey et al.(2016)]{Ca16} Casey, A.~R., Hawkins, K., Hogg,
  D.~W., et al.\ 2016, arXiv:1609.02914

\bibitem[Dehnen(2000)]{De00} Dehnen, W.\ 2000, \aj, 119, 800

\bibitem[Freeman(1993)]{Fr93} Freeman, K.~C.\ 1993, in Smith, G.,
  Brodie, J., eds, ASP Conf. Ser. Vol. 48, The Globular
  Clusters-galaxy Connection. Astron. Soc. Pac., San Francisco, p.608

\bibitem[Gaia Collaboration et al.(2016a)]{GA1} Gaia
  Collaboration, Prusti, T., de Bruijne, J.~H.~J., et al.\ 2016, \aap,
  595, A1

\bibitem[Gaia Collaboration et al.(2016b)]{GA2} Gaia Collaboration,
  Brown, A.~G.~A., Vallenari, A., et al.\ 2016, \aap, 595, A2
  
\bibitem[Green et al.(2015)]{Gr15} Green, G.~M., Schlafly, E.~F.,
  Finkbeiner, D.~P., et al.\ 2015, \apj, 810, 25

\bibitem[Grillmair(2009)]{Gr09} Grillmair, C.~J.\ 2009, \apj, 693, 1118

\bibitem[Helmi et al.(1999)]{He99} Helmi, A., White,
  S.~D.~M., de Zeeuw, P.~T., \& Zhao, H.\ 1999, \nat, 402, 53

\bibitem[Helmi et al.(2017)]{He17} Helmi, A.,
  Veljanoski, J., Breddels, M.~A., Tian, H., \& Sales, L.~V.\ 2017, \aap, 598, A58

\bibitem[H{\o}g et al.(2000)]{Ho00} H{\o}g, E., Fabricius, C.,
  Makarov, V.~V., et al.\ 2000, \aap, 355, L27

\bibitem[Ibata et al.(1994)]{Ib94} Ibata, R.~A.,
  Gilmore, G., \& Irwin, M.~J.\ 1994, \nat, 370, 194

\bibitem[Kunder et al.(2017)]{Ku17} Kunder, A., Kordopatis, G., Steinmetz, M., et al.\ 2017, \aj, 153, 75 

\bibitem[Kordopatis et al.(2013)]{Ko13} Kordopatis, G., Gilmore, G., Steinmetz, M., et al.\ 2013, \aj, 146, 134

\bibitem[Luo et al.(2015)]{Lu15} Luo, A.-L., Zhao,
  Y.-H., Zhao, G., et al.\ 2015, Research in Astronomy and
  Astrophysics, 15, 1095

\bibitem[Majewski et al.(2000)]{Ma00} Majewski, S.~R., Patterson, R.~J., Dinescu, D.~I., et al.\ 2000, in Noels, A., Magain. P., Caro, D., Jehin, E., Parmentier, G., Thoul, A.~A., eds, Proc. 35th Liege Int. Astrophys. Colloq., The Galactic Halo : From Globular Cluster to Field Stars. Institut d'Astrophysique et de Geophysique, Liege, Belgium, p.619

\bibitem[Newberg \& Carlin(2015)]{Ne15} Newberg H.~J., Carlin
  J.~L. 2015, Tidal Streams in the Local Group and Beyond, Springer,
  New York

\bibitem[Smith et al.(2009)]{Sm09} Smith, M.~C., Evans, N.~W.,
  Belokurov, V., et al.\ 2009, \mnras, 399, 1223


\bibitem[Villanova et al.(2007)]{Vi07} Villanova, S.,Piotto, G., King, I.~R., et al.\ 2007, \apj, 663, 296

\bibitem[Villanova et al.(2014)]{Vi14} Villanova, S., Geisler, D.,
  Gratton, R.~G., \& Cassisi, S.\ 2014, \apj, 791, 107

\end{thebibliography}
\end{document}